\documentclass[a4,10pt]{article}
\usepackage{amsmath}
\usepackage{amsfonts}
\usepackage{graphicx}

\input{tcilatex}

\begin{document}

\title{Soliton solutions, B\"{a}cklund transformation and Lax pair for
coupled Burgers system via Bell polynomials}
\author{ \"{O}mer \"{U}nsal\thanks{{\footnotesize Corresponding Author.\ \ \
E-mail address: ounsal@ogu.edu.tr}}, Filiz Ta\c{s}can \\
{\footnotesize Eski\c{s}ehir Osmangazi University, Art-Science Faculty,
Department of Mathematics-Computer, Eski\c{s}ehir-T\"{U}RK\.{I}YE}\\
\\
{\footnotesize Email : ounsal@ogu.edu.tr; ftascan@ogu.edu.tr }}
\maketitle

\begin{abstract}
In this work, we apply the binary Bell polynomial approach to coupled
Burgers system. In other words, we investigate possible integrability of
referred system. Bilinear form and soliton solutions are obtained, some
figures related to these solutions are given. We also get B\"{a}cklund
transformations in both binary Bell polynomial form and bilinear form. Based
on the B\"{a}cklund transformation, Lax pair is obtained. Namely, this is a
study in which integrabilitiy of coupled burgers system is investigated.

\bigskip


\textbf{PACS (2010) :}{\small \ 02.30 Jr, 02.30 lk, 05.45 Yv, 11.30 -j}.
\end{abstract}

\section{Introduction}

Nonlinear evolution equations (NLEEs) and systems of NLEEs are known to
describe a wide variety of phenomena in physics, engineering, applied
mathematics, chemistry, biology. In this manner, integrability of NLEEs and
systems of NLEEs has been more important and topic of many papers. During
the past decades, many scientific workers have presented papers which
includes study of complete integrability of NLEEs. For instance, the Darboux
transformation method \cite{matveev23}, Painleve analysis method \cite{li76}%
, inverse scattering method \cite{ablowitz49}, Hirota bilinear method \cite%
{hirota33}, B\"{a}cklund transformation method \cite{miura88}. Hirota
bilinear method is based on the construction of a bilinear form of given
NLEE. Lambert and his co-workers proposed an useful approach to study the
bilinear form, bilinear B\"{a}cklund transformation, Lax pair, Darboux
transformation by using Bell polynomials \cite{gilson,lambert}. The Bell
polynomial approach is very practical to characterize the bilinearizable
equations. Fan \cite{fan78} generalized this method to search the complete
integrability of the nonisospectral and the variable-coefficient equations
and proposed a procedure to obtain the infinite conservation laws of the
NLEEs. Lambert and Springael constructed Bell polynomial approach for
systems of NLEEs and NLEEs which can be transformed to a system of NLEEs.
They also gave some applications \cite{lambert8}. Then, some authors
presented more applications \cite{sunya,yi,wang46,wang9,yan89}.

This paper is organized as follows: In Section 2, we briefly present some
notations related to Bell polynomials used in the literature. In Section 3,
bilinear formalism and soliton solutions are obtained, some figures
correspond to the soliton solutions are given. In section 4, B\"{a}cklund
transformation and Lax pair are costructed. In section 4, some conclusions
are given.

\section{Bell polynomials and binary ones}

In this section, we give the basic definitions and fundamentals of
multi-dimensional Bell polynomials and binary Bell polynomials. For details,
please refer to \cite%
{wang9,yan89,bell1,lambert3,lambert4,lambert5,lambert6,pempinelli,jiang}.
Consider the function $f=f(x_{1},x_{2},...,x_{l})$ with multi-variables in $%
C^{\infty }.$ The multi-dimensional Bell polynomials are given as%
\begin{equation}
\begin{array}{lll}
Y_{n_{1}x_{1},...,n_{l}x_{l}}(f) & = & 
Y_{n_{1},...,n_{l}}(f_{r_{1}x_{1},...,r_{l}x_{l}}) \\ 
& = & e^{-f}\partial _{x_{1}}^{n_{1}}...\partial _{x_{l}}^{n_{l}}e^{f},%
\end{array}
\tag{1}
\end{equation}%
where%
\begin{equation*}
\begin{array}{c}
f_{r_{1}x_{1},...,r_{l}x_{l}}=\partial _{x_{1}}^{r_{1}}...\partial
_{x_{l}}^{r_{l}}f, \\ 
\\ 
r_{1}=0,...,n_{1};...;r_{l}=0,...,n_{l}.%
\end{array}%
\end{equation*}%
For the simplest case $f=f(x),$ the one-dimensional Bell polynomials are
expressed as%
\begin{equation}
\begin{array}{l}
Y_{1}=f_{x},Y_{2}=f_{2x}+f_{x}^{2}, \\ 
Y_{3}=f_{3x}+3f_{x}f_{2x}+f_{x}^{3},....%
\end{array}
\tag{2}
\end{equation}%
If function has two variables, namely for the $f=f(x,t),$ the associated
two-dimensional Bell polynomials are%
\begin{equation}
\begin{array}{lll}
Y_{x,t} & = & f_{x,t}+f_{x}f_{t}, \\ 
Y_{2x,t} & = & f_{2x,t}+f_{2x}f_{t}+2f_{x,t}f_{x}+f_{x}^{2}f_{t},....%
\end{array}
\tag{3}
\end{equation}%
The multi-dimensional binary Bell polynomials take the following forms%
\begin{equation}
\mathcal{Y}_{n_{1}x_{1},...,n_{l}x_{l}}(v,w)=Y_{n_{1},...,n_{l}}(f)  \tag{4}
\end{equation}%
for%
\begin{equation*}
f_{r_{1}x_{1},...,r_{l}x_{l}}=\left\{ 
\begin{array}{l}
v_{r_{1}x_{1},...,r_{l}x_{l}},\text{ when }r_{1}+...+r_{l}\text{ is odd} \\ 
w_{r_{1}x_{1},...,r_{l}x_{l}},\text{ when }r_{1}+...+r_{l}\text{ is even.}%
\end{array}%
\right.
\end{equation*}%
The first few lowest order binary Bell polynomials read%
\begin{equation}
\begin{array}{l}
\mathcal{Y}_{x}(v)=v_{x},\text{ }\mathcal{Y}_{2x}(v,w)=w_{2x}+v_{x}^{2}, \\ 
\mathcal{Y}_{x,t}(v,w)=w_{x,t}+v_{x}v_{t}, \\ 
\mathcal{Y}_{3x}(v,w)=v_{3x}+3v_{x}w_{2x}+v_{x}^{3}, \\ 
\mathcal{Y}%
_{4x}(v,w)=w_{4x}+3w_{2x}^{2}+4v_{x}v_{3x}+6v_{x}^{2}w_{2x}+v_{x}^{4},....%
\end{array}
\tag{5}
\end{equation}%
The link between the $\mathcal{Y}$-pollynomials and the standard Hirota
bilinear formula $D_{x_{1}}^{n_{1}}...D_{x_{l}}^{n_{l}}F.G$ can be expressed
by the following identity:%
\begin{equation}
\mathcal{Y}_{n_{1}x_{1},...,n_{l}x_{l}}(v=\ln F/G,w=\ln FG)=\left( FG\right)
^{-1}D_{x_{1}}^{n_{1}}...D_{x_{l}}^{n_{l}}F.G,  \tag{6}
\end{equation}%
where $n_{1}+n_{2}+...+n_{l}\geq 1,$ and operators $D_{x_{1}},...,D_{x_{l}}$
are the classical Hirota's bilinear operators defined by%
\begin{equation*}
D_{x_{1}}^{n_{1}}...D_{x_{l}}^{n_{l}}F.G=\left( \partial _{x_{1}}-\partial
_{x_{1}^{\prime }}\right) ^{n_{1}}...\left( \partial _{x_{l}}-\partial
_{x_{l}^{\prime }}\right) ^{n_{l}}F(x_{1},...,x_{l})\times G(x_{1}^{\prime
},...,x_{l}^{\prime })\mid _{x_{1}^{\prime }=x_{1},...,x_{l}^{\prime
}=x_{l}}.
\end{equation*}

Furthermore, the relationship between the binary Bell polynomials and Lax
Pair by use of the Hopf-Cole transformation is given by the expression%
\begin{equation}
\mathcal{Y}_{n_{1}x_{1},n_{2}x_{2}}(v=\ln \psi ,w=v+Q)=\psi
^{-1}\sum\limits_{r=0}^{n_{1}}\sum\limits_{s=0}^{n_{2}}\binom{n_{1}}{r}%
\binom{n_{2}}{s}\mathcal{Y}_{rx_{1},sx_{2}}\left( 0,Q\right) \partial
_{x_{1}}^{n_{1}-r}\partial _{x_{2}}^{n_{2}-s}\psi ,  \tag{7}
\end{equation}%
where $\psi $ and $Q$ are both the functions of $x_{1}$ and $x_{2}.$

\section{Binary Bell polynomial form and soliton solutions}

Now we consider the coupled Burgers system \cite{wang33}%
\begin{equation}
u_{t}-2uu_{x}-v_{xx}=0  \tag{8a}
\end{equation}%
\begin{equation}
v_{yt}-u_{xxy}-2uv_{xy}-2u_{x}v_{y}=0.  \tag{8b}
\end{equation}%
Wang et al \cite{wang33} obtained infinitely many generalized symmetries of
system (8). We introduce two dimensionless fields $p$ and $q$ which are the
functions of $x,y$ and $t.$ Taking the transformation%
\begin{equation}
u=p_{x}\text{ },\text{ }v=q_{x}  \tag{9}
\end{equation}%
into account and integrating the resulting equation once with respect to $x$%
, we get the Eq.(8a) as%
\begin{equation}
p_{t}-p_{x}^{2}-q_{xx}=0.  \tag{10}
\end{equation}%
Substituting (9) into Eq.(8b), using Eq.(10) and integrating the resulting
equation once with respect to $x,$ we obtain%
\begin{equation}
q_{yt}+p_{y}p_{t}-p_{y}\left( p_{x}^{2}+q_{xx}\right)
-p_{xxy}-2p_{x}q_{xy}=0.  \tag{11}
\end{equation}%
From Eqs.(10) and (11) we find binary Bell polynomial form of the system (8)
as%
\begin{equation}
\mathcal{Y}_{t}(p,q)-\mathcal{Y}_{xx}(p,q)=0  \tag{12a}
\end{equation}%
\begin{equation}
\mathcal{Y}_{yt}(p,q)-\mathcal{Y}_{xxy}(p,q)=0.  \tag{12b}
\end{equation}%
Through expression (6) and setting that $p=\ln \left( f/g\right) $ $,$ $%
q=\ln \left( fg\right) $ we get bilinear form of system (8) as%
\begin{equation}
\left[ D_{t}-D_{x}^{2}\right] f.g=0  \tag{13a}
\end{equation}%
\begin{equation}
\left[ D_{y}D_{t}-D_{x}^{2}D_{y}\right] f.g=0  \tag{13b}
\end{equation}%
where $f$ and $g$ are the functions of $x,y$ and $t.$

As follows, the Eqs.(13) can be easily solved to obtain the multi-soliton
solutions of system (8) by using the Hirota's bilinear method.

Let $f$ and $g$ be functions in the form%
\begin{equation}
f=\epsilon f_{1}+\epsilon ^{2}f_{2}+...  \tag{14a}
\end{equation}%
\begin{equation}
g=1+\epsilon g_{1}+\epsilon ^{2}g_{2}+....  \tag{14b}
\end{equation}%
Substituting expressions (14) into Eqs. (13) and collecting the coefficients
of the same power of $\epsilon ,$ we have 
\begin{equation}
\epsilon ^{1}:\text{ }\left[ D_{t}-D_{x}^{2}\right] \left( f_{1}.1\right) =0
\tag{15a}
\end{equation}%
\begin{equation}
\left[ D_{y}D_{t}-D_{x}^{2}D_{y}\right] \left( f_{1}.1\right) =0  \tag{15b}
\end{equation}%
\begin{equation}
\epsilon ^{2}:\text{ }\left[ D_{t}-D_{x}^{2}\right] \left(
f_{1}.g_{1}+f_{2}.1\right) =0  \tag{16a}
\end{equation}%
\begin{equation}
\left[ D_{y}D_{t}-D_{x}^{2}D_{y}\right] \left( f_{1}.g_{1}+f_{2}.1\right) =0
\tag{16b}
\end{equation}%
\begin{equation}
\epsilon ^{3}:\text{ }\left[ D_{t}-D_{x}^{2}\right] \left(
f_{1}.g_{2}+f_{2}.g_{1}+f_{3}.1\right) =0  \tag{17a}
\end{equation}%
\begin{equation}
\left[ D_{y}D_{t}-D_{x}^{2}D_{y}\right] \left(
f_{1}.g_{2}+f_{2}.g_{1}+f_{3}.1\right) =0  \tag{17b}
\end{equation}%
\begin{equation*}
\vdots
\end{equation*}

(i) One-Soliton Solution

In order to find one soliton solution of system (8), we take%
\begin{equation}
\begin{array}{c}
f_{1}=ae^{\xi },\text{ }g_{1}=be^{\xi },\text{ }\xi =kx+my+wt, \\ 
\\ 
f_{i}(x,y,t)=0,g_{i}(x,y,t)=0\text{ }i=2,3,4,...%
\end{array}
\tag{18}
\end{equation}%
where $a,b,k$ and $m$ are all arbitrary nonzero real constants. $w$ is a
real constant to be determined. Solving (15)-(17) which found by
substituting expressions (14) into (13), we get $w=k^{2}.$ So the one
soliton solutions for system (8) is obtained in the form%
\begin{equation}
u=\frac{k}{1+be^{kx+my+k^{2}t}}  \tag{19a}
\end{equation}%
\begin{equation}
v=\frac{k\left( 1+2be^{kx+my+k^{2}t}\right) }{1+be^{kx+my+k^{2}t}}  \tag{19b}
\end{equation}

(i) Two-Soliton Solution

Similarly, we look for the two-soliton solution by choosing%
\begin{equation}
\begin{array}{c}
f_{1}=a_{1}e^{\xi _{1}}+a_{2}e^{\xi _{2}},\text{ }g_{1}=b_{1}e^{\xi
_{1}}+b_{2}e^{\xi _{2}}, \\ 
\\ 
f_{2}=a_{3}e^{\xi _{1}+\xi _{2}},\text{ }g_{2}=b_{3}e^{\xi _{1}+\xi _{2}},%
\text{ }\xi _{i}=k_{i}x+m_{i}y+w_{i}t,\text{ }i=1,2%
\end{array}
\tag{20}
\end{equation}%
where $a_{1},a_{2},a_{3},b_{1},b_{2},b_{3},k_{1},k_{2}$ are all arbitrary
nonzero constants. We truncate the expressions (14) to $f_{2}(x,y,t)$ and $%
g_{2}(x,y,t),$ respectively. Then, corresponding two-soliton solution is%
\begin{equation}
u=\left[ \ln \left( \frac{a_{1}e^{\xi _{1}}+a_{2}e^{\xi _{2}}+a_{3}e^{\xi
_{1}+\xi _{2}}}{1+b_{1}e^{\xi _{1}}+b_{2}e^{\xi _{2}}+b_{3}e^{\xi _{1}+\xi
_{2}}}\right) \right] _{x}  \tag{21a}
\end{equation}%
\begin{equation}
v=\left[ \ln \left( \left( a_{1}e^{\xi _{1}}+a_{2}e^{\xi _{2}}+a_{3}e^{\xi
_{1}+\xi _{2}}\right) \left( 1+b_{1}e^{\xi _{1}}+b_{2}e^{\xi
_{2}}+b_{3}e^{\xi _{1}+\xi _{2}}\right) \right) \right] _{x}  \tag{21b}
\end{equation}%
where%
\begin{equation}
a_{1}=\frac{m_{1}k_{1}a_{3}}{b_{2}(k_{1}-k_{2})\left( m_{1}-m_{2}\right) }%
\text{ },\text{ }b_{1}=\frac{m_{2}k_{2}a_{3}}{a_{2}(k_{1}-k_{2})\left(
m_{1}-m_{2}\right) }\text{ },\text{ }w_{i}=k_{i}^{2}.  \tag{22}
\end{equation}%
Some figures which corresponds to soliton solutions found above are given
below. Figs. 1-2 and 3-6 represents (19) and (21) respectively. Figs. 1 and
2 show the motion of the shock waves. Fig. 3 shows head on collision of two
kink type solitary waves. The collision is inelastic, that's why the shape
of waves is changed after the collision. Fig. 5 depicts the interaction of
two bell shaped solitary waves. Waves move emerging into one after the
collision. Then the collision is inelastic. Figs. 4 and 6 show the
interaction of two kink type solitons. The solitons keep their own sahpe and
direction after the interaction. So collision is elastic.

\begin{center}
\begin{tabular}{cc}
\includegraphics[width=2.0in]{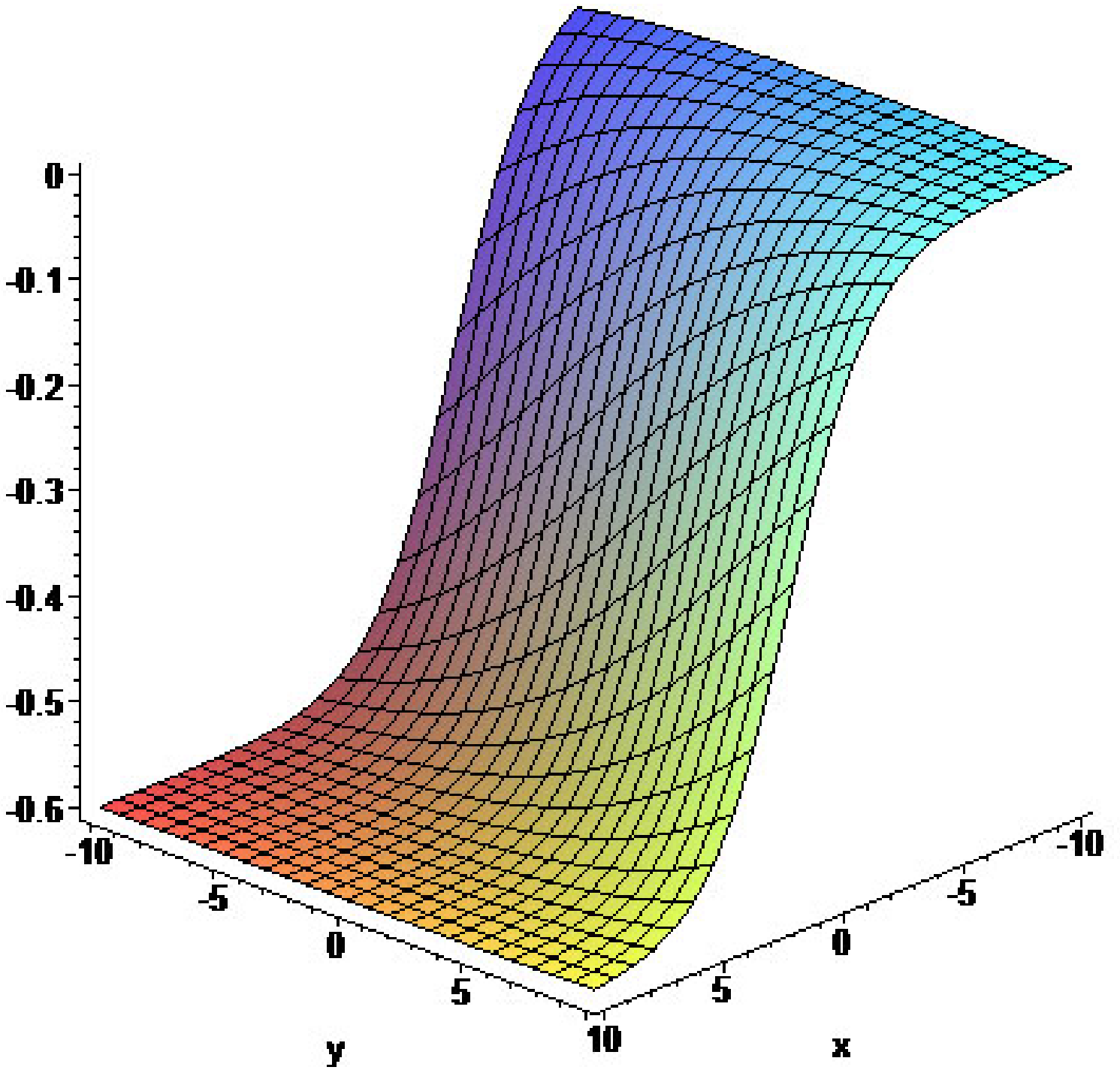} & \includegraphics[width=2.0in]{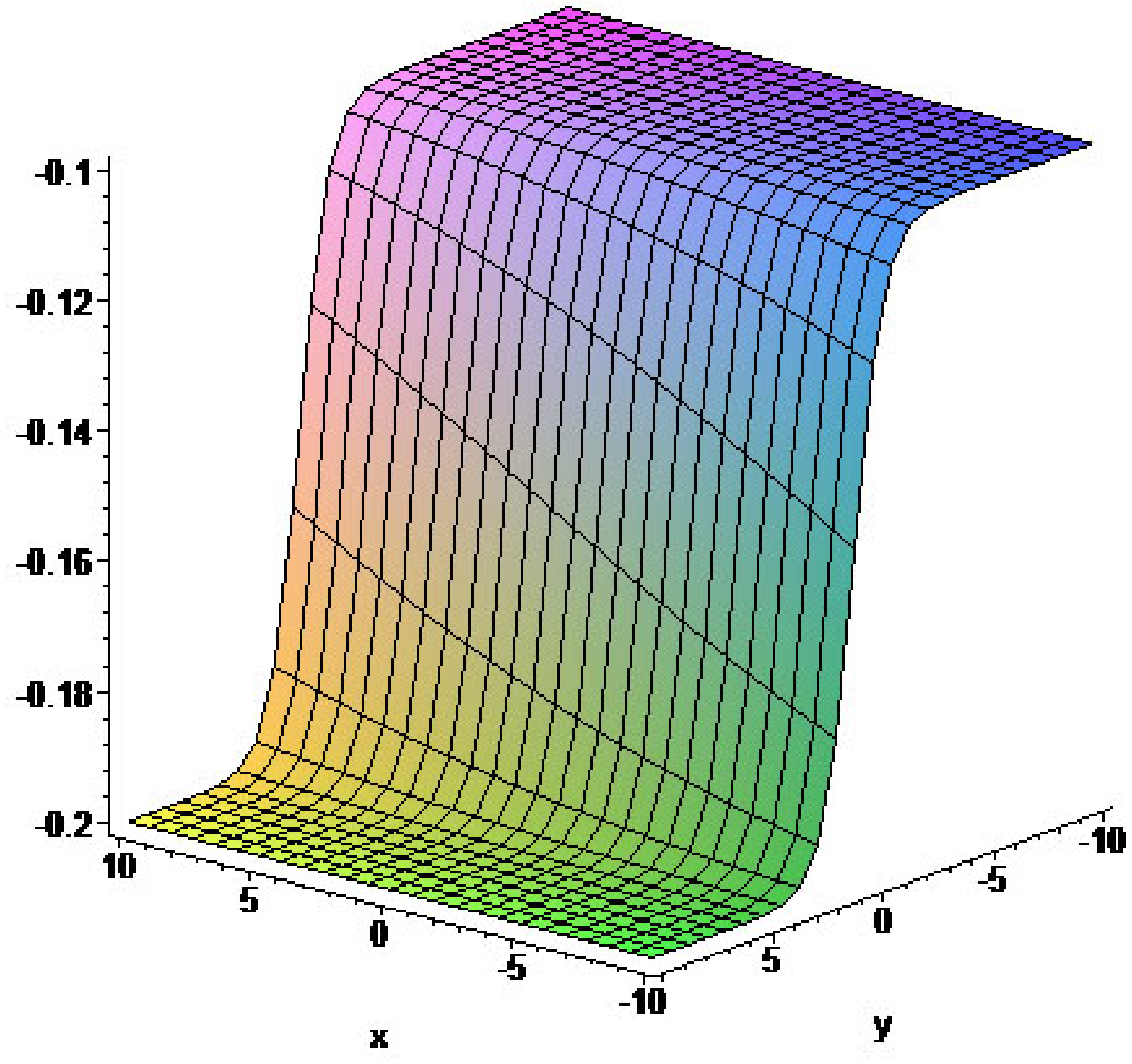} \\ 
u:Fig.\text{ }1.\text{ }$t=0,k=-0.6,m=0.2,b=0.4$ & v:Fig.\text{ }2.\text{ }$%
t=0,k=-0.1,m=1.5,b=0.1$%
\end{tabular}

\begin{tabular}{cc}
\includegraphics[width=2.0in]{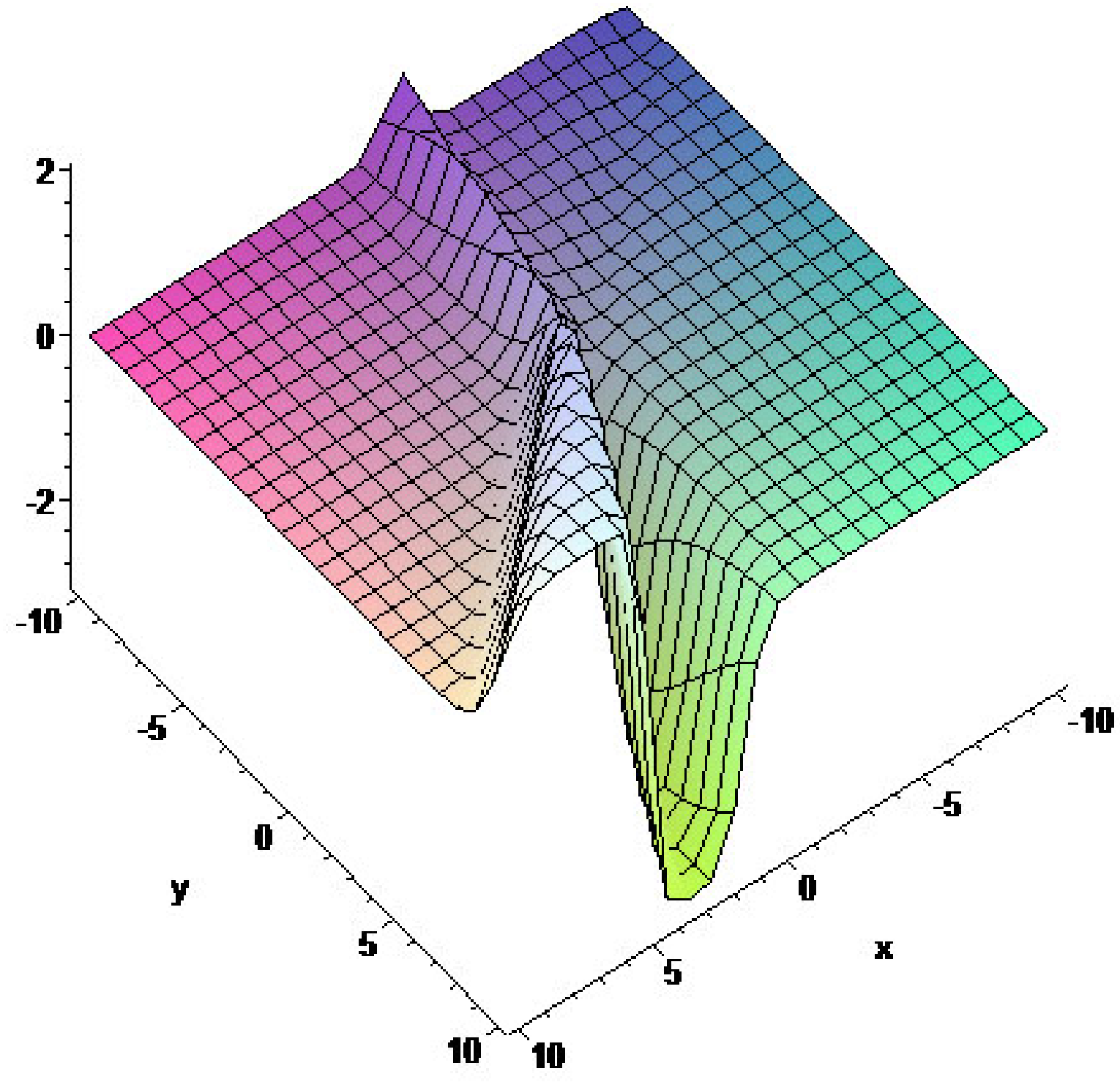} & \includegraphics[width=2.0in]{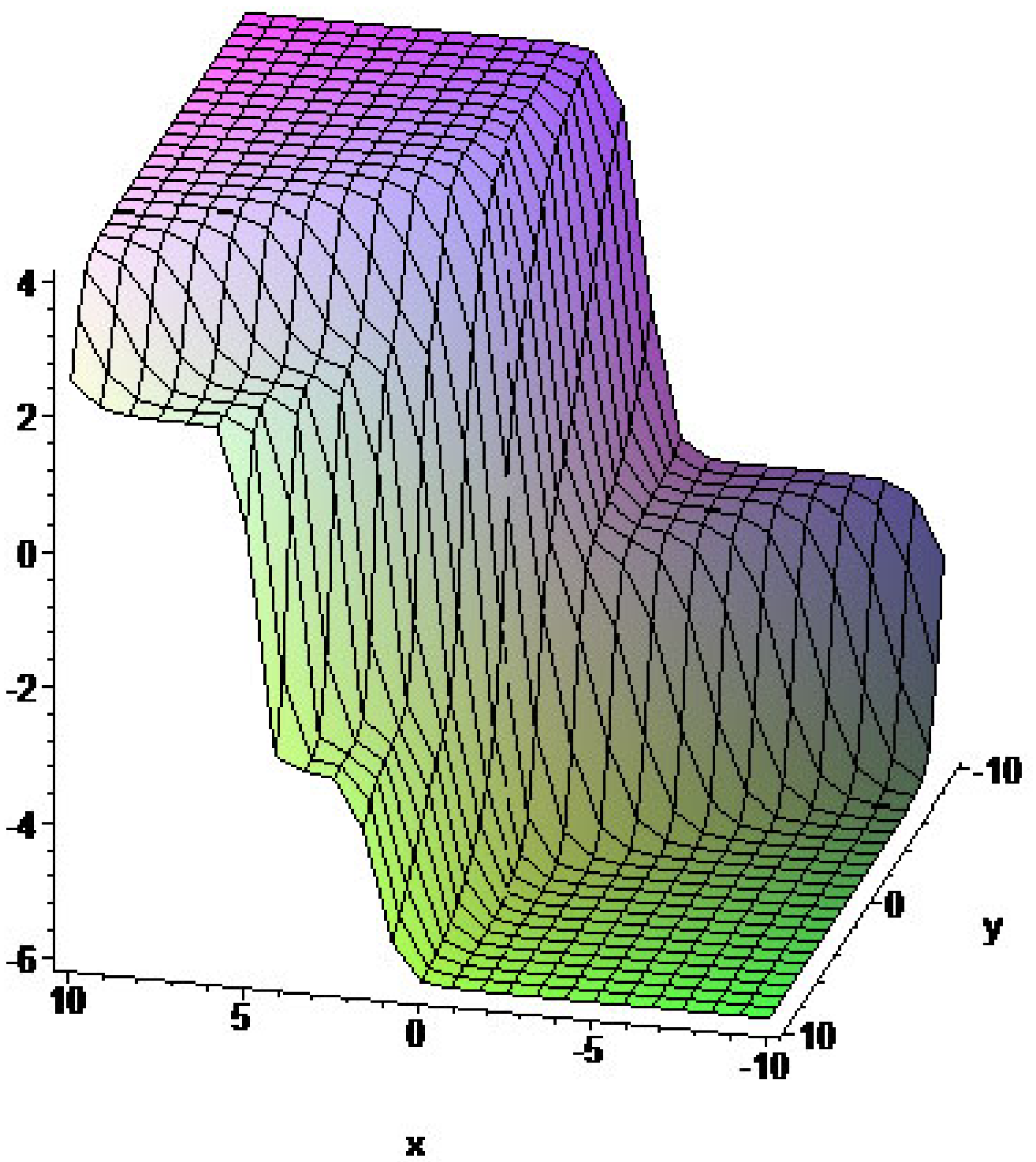} \\ 
u:Fig.\text{ }3.\text{ }$t=0,k_{1}=-3,k_{2}$=2, & v:Fig. 4. $%
t=0,k_{1}=-3,k_{2}=2$, \\ 
$m_{1}=0.5,m_{2}=-2,a_{2}=-0.2,a_{3}=-0.2,$ & $%
m_{1}=0.5,m_{2}=-2,a_{2}=-0.2,a_{3}=-0.2,$ \\ 
$b_{2}=0.3,b_{3}=1$ & $b_{2}=0.3,b_{3}=1$%
\end{tabular}

\begin{tabular}{cc}
\includegraphics[width=2.0in]{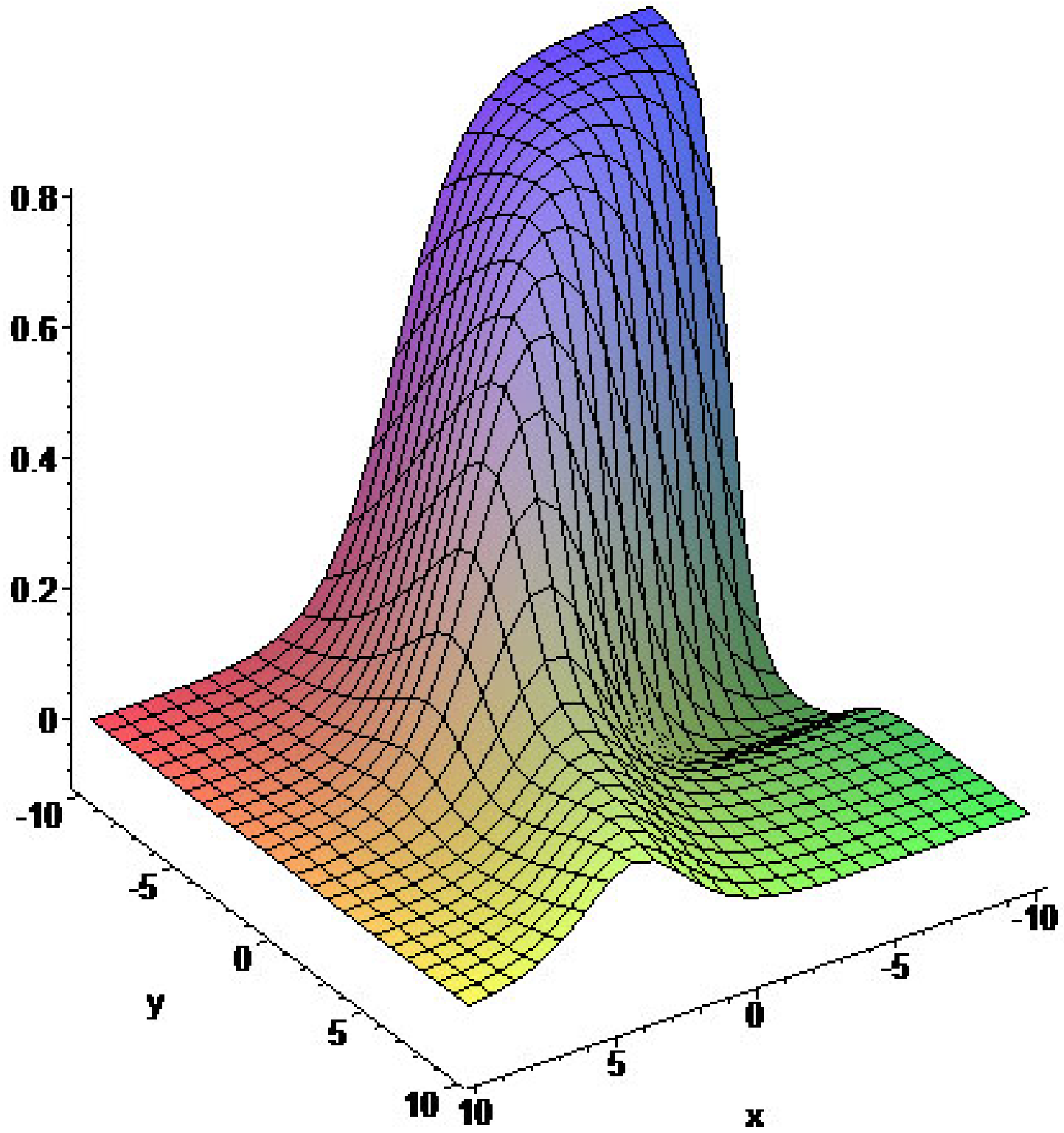} & \includegraphics[width=2.0in]{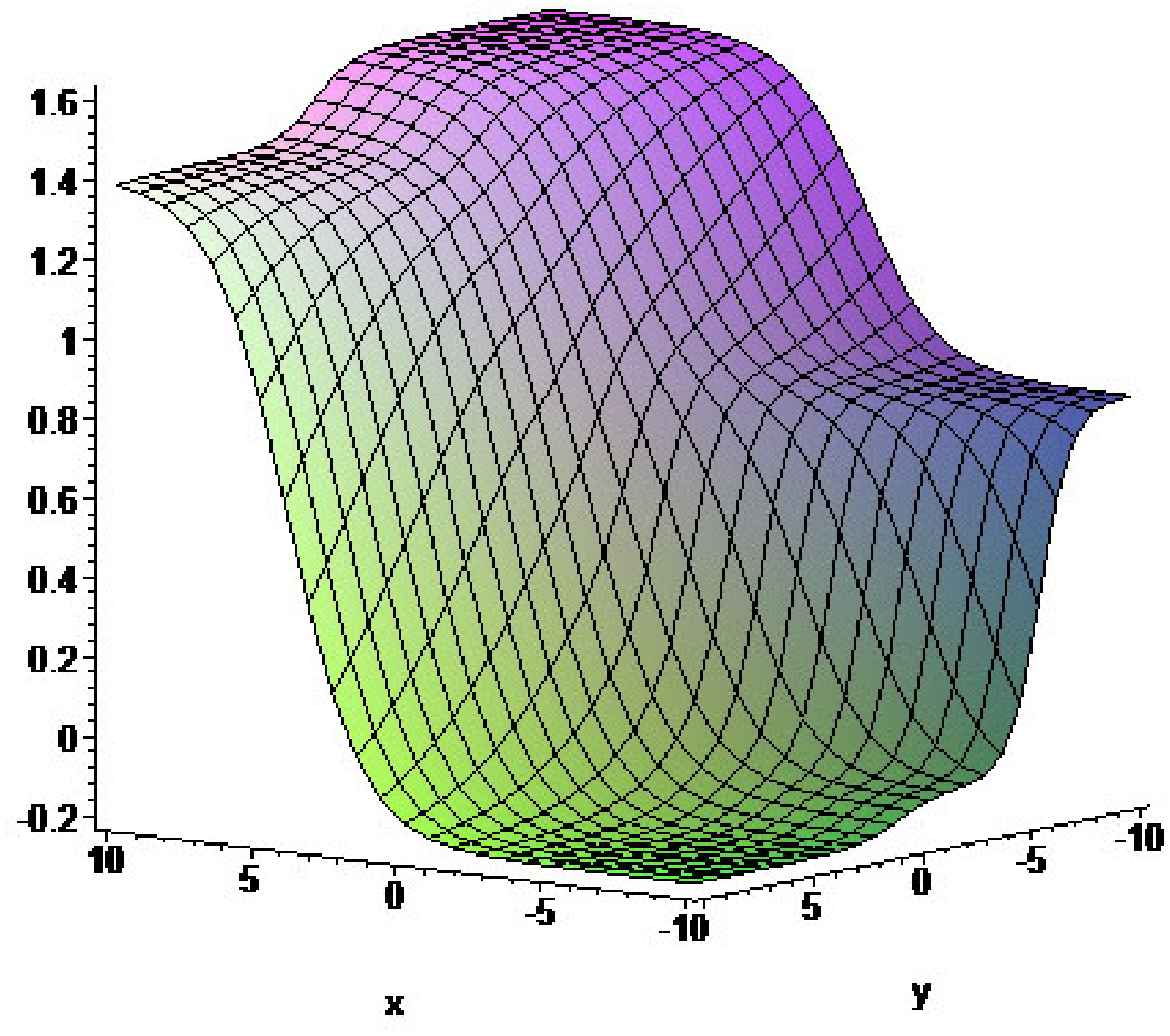} \\ 
u:Fig.\text{ }5.\text{ }$t=0,k_{1}=-0.1,k_{2}=$0.8, & v:Fig. 6. $%
t=0,k_{1}=-0.1,k_{2}=0.8,$ \\ 
$m_{1}=1.2,m_{2}=-0.3,a_{2}=-0.2,a_{3}=-0.2,$ & $%
m_{1}=1.2,m_{2}=-0.3,a_{2}=-0.2,a_{3}=-0.2,$ \\ 
$b_{2}=0.1,b_{3}=1$ & $b_{2}=0.1,b_{3}=1$%
\end{tabular}
\end{center}

\section{B\"{a}cklund transformation and Lax Pair}

In this section, we get the bilinear B\"{a}cklund transformation in both the
binary Bell polynomial and bilinear forms. We also obtain Lax pair of system
(8). We take%
\begin{equation}
P_{1}=\left[ \mathcal{Y}_{t}(p^{\prime },q^{\prime })-\mathcal{Y}%
_{xx}(p^{\prime },q^{\prime })\right] -\left[ \mathcal{Y}_{t}(p,q)-\mathcal{Y%
}_{xx}(p,q)\right]  \tag{23}
\end{equation}%
\begin{equation}
P_{2}=\left[ \mathcal{Y}_{yt}(p^{\prime },q^{\prime })-\mathcal{Y}%
_{xxy}(p^{\prime },q^{\prime })\right] -\left[ \mathcal{Y}_{yt}(p,q)-%
\mathcal{Y}_{xxy}(p,q)\right]  \tag{24}
\end{equation}%
where $\left( p^{\prime },q^{\prime }\right) $ and $\left( p,q\right) $
satisfy the Eqs. (12). Supposing that $p^{\prime }=\ln \left( f^{\prime
}/g^{\prime }\right) $ and $q^{\prime }=\ln \left( f^{\prime }g^{\prime
}\right) $, using the relations%
\begin{equation}
\begin{array}{cccc}
v_{1}=\ln \left( g^{\prime }/g\right) , & v_{2}=\ln \left( f^{\prime
}/f\right) , & v_{3}=\ln \left( f^{\prime }/g\right) , & v_{4}=\ln \left(
g^{\prime }/f\right) , \\ 
w_{1}=\ln \left( g^{\prime }g\right) , & w_{2}=\ln \left( f^{\prime
}f\right) , & w_{3}=\ln \left( f^{\prime }g\right) , & w_{4}=\ln \left(
g^{\prime }f\right) ,%
\end{array}
\tag{25}
\end{equation}%
\begin{equation}
\begin{array}{ll}
p^{\prime }-p=v_{2}-v_{1}=w_{3}-w_{4} & p^{\prime }+p=v_{3}-v_{4}=w_{2}-w_{1}
\\ 
q^{\prime }-q=v_{1}+v_{2}=v_{3}+v_{4} & q^{\prime }+q=w_{1}+w_{2}=w_{3}+w_{4}%
\end{array}
\tag{26}
\end{equation}%
we obtain the binary Bell polynomial form of B\"{a}cklund transformation for
the system (8)%
\begin{equation}
\begin{array}{l}
\mathcal{Y}_{t}(v_{1},w_{1})+\mathcal{Y}_{xx}(v_{1},w_{1})=0 \\ 
\mathcal{Y}_{t}(v_{2},w_{2})+\mathcal{Y}_{xx}(v_{2},w_{2})=0 \\ 
\mathcal{Y}_{x}(v_{3},w_{3})-\mu e^{v_{1}-v_{2}}=0 \\ 
\mathcal{Y}_{xy}(v_{3},w_{3})-d\mathcal{Y}_{x}(v_{3},w_{3})+\mu
e^{v_{1}-v_{2}}\mathcal{Y}_{y}(v_{4},w_{4})=0%
\end{array}
\tag{27}
\end{equation}%
where $\mu $ and $d$ are arbitrary real constants. With the aid of (6), we
can also express B\"{a}cklund transformation in the bilinear form as follows:%
\begin{equation}
\begin{array}{l}
\left[ D_{t}+D_{x}^{2}\right] g^{\prime }.g=0 \\ 
\left[ D_{t}+D_{x}^{2}\right] f^{\prime }.f=0 \\ 
D_{x}f^{\prime }.g-\mu f^{\prime }g=0 \\ 
\left[ D_{x}D_{y}-dD_{x}\right] f^{\prime }.g+\mu D_{y}g^{\prime }.f=0.%
\end{array}
\tag{28}
\end{equation}%
As a result of the relations (25)-(26), relations%
\begin{equation}
\begin{array}{cccc}
v_{2}=v_{3}-p, & w_{2}=w_{3}+p, & v_{4}=v_{1}-p, & w_{4}=w_{1}+p%
\end{array}
\tag{29}
\end{equation}%
can be written \cite{lambert8}. Taking that%
\begin{equation}
\begin{array}{ccc}
w_{i}=v_{i}+Q_{i}, & v_{i}=\ln \psi _{i}, & \left( i=1,3\right)%
\end{array}
\tag{30}
\end{equation}%
we can use the relations%
\begin{equation}
Q=w_{1}-v_{1}=w_{3}-v_{3}=Q_{1}=Q_{3}.  \tag{31}
\end{equation}%
Using (29)-(31), we find the Lax pair for the system (8) in the form%
\begin{equation}
\begin{array}{c}
\psi _{1,y}=\frac{d}{2}\psi _{1}-\frac{e^{-p}\left( q_{xy}-p_{xy}\right) }{%
2\mu }\psi _{3} \\ 
\\ 
\psi _{1,t}=\left( p_{xx}-q_{xx}\right) \psi _{1}-\psi _{1,xx} \\ 
\\ 
\psi _{3,x}=\mu e^{p}\psi _{1} \\ 
\\ 
\psi _{3,t}=\mu p_{x}e^{p}\psi _{1}-\mu e^{p}\psi _{1,x}.%
\end{array}
\tag{32}
\end{equation}%
Compatibility conditions $\psi _{1,yt}=\psi _{1,ty}$ $,$ $\psi _{3,xt}=\psi
_{3,tx}$ can easily be verified.

\section{\ Conclusion}

In this study, we have investigated the coupled Burgers system with the aid
of Bell polynomials. In this sense, we have obtained Bilinear form. Using
bilinear form we have gotten soliton solutions and discussed these solutions
analytically. \ Some figures of one-soliton and two-soliton solutions have
been given. B\"{a}cklund transformations in both binary Bell polynomial form
and bilinear form have been given. Then, Lax pair has been constructed. We
hope that the present findings may be useful in further works.


\begin{thebibliography}{99}
\bibitem{matveev23} Matveev, V.B., Salle, M.A., Darboux transformation and
solutions, Springer, Berlin (1980).

\bibitem{li76} Li, X.N., Wei, G.M., Liang, Y.Q., Painleve analysis and new
analytic solutions for variable-coefficient Kadomtsev-Petviashvili equation
with symbolic computation, Appl. MAth. Comput. 216, (2010) 3568.

\bibitem{ablowitz49} Ablowitz, M.J., Clarkson, P.A., Solitonsi nonlinear
evolution equations and inverse scattering, Cambridge University Press,
Cambridge (1991).

\bibitem{hirota33} Wazwaz, A.M., Multiple-soliton solutions for the KP
equation by Hirota's bilinear method and by the tanh--coth method, Applied
Mathematics and Computation, 190,1 (2007) 633.

\bibitem{miura88} Miura, M.R., Backlund Transformation, Springer-Verlag,
Berlin, (1978).

\bibitem{gilson} Gilson, C., Lambert, F., Nimmo, J., Willox, R., On the
combinatorics of the Hirota D-operators, Proc. R. Soc. Lond. A, 452, (1996)
223.

\bibitem{lambert} Lambert, F., Springael, J., Soliton equations and simple
combinatorics, Acta. Appl. Math., 102, (2008) 147.

\bibitem{fan78} Fan, E., The integrability of nonisospectral and
variable-coefficient KdV equation with binary Bell polynomials, Physics
Letters A, 375, (2011) 493.

\bibitem{lambert8} Lambert, F., Springael, J., On a direct procedure for the
disclosure of Lax pairs and B\"{a}cklund transformations, Chaos, Solitons
and Fractals, 12, (2001) 2821.

\bibitem{sunya} Sun, Y., Tian, B., Sun, W.R., Jiang, Yan., Wang, Y.P.,
Huang, Z.R., B\"{a}cklund transformation and N-soliton solutions for a
(2+1)-dimensional nonlinear evolution equation in nonlinear water waves.

\bibitem{yi} Qin, Y., Gao, Y.T., Yu, X., Meng, G.Q., Bell polynomial
approach and N-Soliton Solutions for a Coupled KdV-mKdV System, Commun.
Theor. Phys., 58, (2012) 73.

\bibitem{wang46} Wang, Y.F., Tian, B., Wang, P., Li, M., Jiang, Y.,
Bell-polynomial approach and soliton solutions for the Zhiber-Shabat
equation and (2+1)-dimensional Gardner equation with symbolic computation,
Nonlinear Dyn., 69, (2012) 2031.

\bibitem{wang9} Wang, P., Tian, B., Liu, W.J., L\"{u}, X., Jiang, Y., Lax
pair, B\"{a}cklund transformation and multi-soliton solutions for the
Boussinesq-Burgers equations from shallow water waves, Applied Mathematics
and Computation, 218, (2011) 1726.

\bibitem{yan89} Jiang, Y., Tian, B., Liu, W.J., Sun, K., Li, M., Soliton
solutions and integrability for the generalized variable-coefficient
extended Korteweg-de Vries equation in fluids, Applied Mathematics Letters,
26, (2013) 402.

\bibitem{bell1} Bell, E.T., Exponential polynomials, Ann. Math. 35, (1934)
258.

\bibitem{lambert3} Lambert, F., Loris, I., Springael, J., Willox, R., On
modified NLS, Kaup and NLBqequations: differential transformations and
bilinearization, J. Phys. A: Math. Gen. 27, (1994) 8705..

\bibitem{lambert4} Lambert, F., Springael, J., Construction of B\"{a}cklund
transformations with binary Bell polynomials, J. Phys. Soc. Jpn., 66, (1997)
2211.

\bibitem{lambert5} Lambert, F., Loris, I., Springael, J., Classical Darboux
transformations and the KP hierarchy, Inverse Problems, 17, (2001) 1067.

\bibitem{lambert6} Lambert, F., Leble, S., Springael, J., Binary Bell
polynomials and Darboux covariant Lax pairs, Glasg. Math. J.,43A, (2001) 53.

\bibitem{pempinelli} Luo, L., New exact solutions and Backlund
transformation for Boiti-Leon-Manna-Pempinelli equation, Physics Letters A,
375, (2011) 1059.

\bibitem{jiang} Jiang, Y., Tian, B., Liu, W.J., Li, M., Wang, P., Sun, K.,
Solitons, Backlund transformation, and Lax pair for the (2+1)-dimensional
Boiti-Leon-Pempinelli equation for the water waves, Journal of Mathematical
Physics, 51, (2010) 093519.

\bibitem{wang33} Wang, Y.J., Liang, Z.F., Tang, X.Y., Infinitely many
generalized symmetries and Painleve analysis of a (2+1)-dimensional Burgers
system, Physica Scripta, 89, (2014) 025201.
\end{thebibliography}
\end{document}